\documentclass[prb,showpacs,altaffilletter,preprint]{revtex4}
\usepackage{graphicx}
\usepackage{dcolumn}
\usepackage{bm}
\usepackage{relsize}

\begin{document}

\title{Field Effect Transistor Based on KTaO$\rm{}_3$ Perovskite}
\author{K.~Ueno}
 \altaffiliation[Electronic mail: ]{kazunori-ueno@aist.go.jp}
 \altaffiliation[Also at: ]{Department of Advanced Materials Science,
	University of Tokyo, Kashiwa, Chiba 277-8581, Japan}
 \affiliation{Correlated Electron Research Center (CERC), National
  Institute of Advanced Industrial Science and Technology (AIST),
  Tsukuba 305-8562, Japan}

\author{I.~H.~Inoue}
 \affiliation{Correlated Electron Research Center (CERC), National
  Institute of Advanced Industrial Science and Technology (AIST),
  Tsukuba 305-8562, Japan}

\author{T.~Yamada}
 \affiliation{Correlated Electron Research Center (CERC), National
  Institute of Advanced Industrial Science and Technology (AIST),
  Tsukuba 305-8562, Japan}

\author{H.~Akoh}
 \affiliation{Correlated Electron Research Center (CERC), National
  Institute of Advanced Industrial Science and Technology (AIST),
  Tsukuba 305-8562, Japan}

\author{Y.~Tokura}
\altaffiliation[Also at: ]{Department of Applied Physics, University
  of Tokyo, Bunkyo-ku, Tokyo 113-8656, Japan}
\affiliation{Correlated Electron Research Center (CERC), National
  Institute of Advanced Industrial Science and Technology (AIST),
  Tsukuba 305-8562, Japan}

\author{H.~Takagi}
 \altaffiliation[Also at: ]{Department of Advanced Materials Science,
	University of Tokyo, Kashiwa, Chiba 277-8581, Japan, and CREST, JST}
 \affiliation{Correlated Electron Research Center (CERC), National
  Institute of Advanced Industrial Science and Technology (AIST),
  Tsukuba 305-8562, Japan}

\begin{abstract}
An n-channel accumulation-type field effect transistor (FET) has been fabricated utilizing a $\rm KTaO_3$ single crystal
as an active element and a sputtered amorphous $\rm Al_2O_3$ film as a gate insulator. The device demonstrated an ON/OFF ratio of $\rm 10^4$ and a field effect mobility of 0.4\,$\rm cm^2/Vs$ at room temperature,
both of which are much better than those of the $\rm SrTiO_3$ FETs reported previously.
The field effect mobility  was almost temperature independent down to 200\,K.
Our results indicate that the $\rm Al_2O_3 / KTaO_3$ interface is worthy of
further investigations as an alternative system of future oxide electronics.
\end{abstract}
\pacs{
85.30.Tv, 
71.30.+h, 
72.80.Ga, 
73.40.Qv 
}

\maketitle

Solid state devices based on transition metal oxides (TMO), especially with perovskite related structure,
are very promising candidates for the next generation electronics due to their rich variety of functions
such as superconductivity, ferroelectricity, and colossal magnetoresistance.
Among those devices, the field effect transistor (FET) is the most fundamental one,
and thus the fabrication of FETs using perovskite-related oxides for conducting channels is a first step in a large movement towards oxide electronics.
Nevertheless, only a small number of perovskite FETs with relatively low mobilities have been reported so far.
\cite{ybco_fet,mott_fet,italy_sto,sto_thesis}
In this letter, we explore a new perovskite-FET with higher mobility characteristics.

The key component of our new device is $\rm KTaO_3$,
an n-type semiconductor with a band gap of 3.8\,eV, which, in single crystalline form, exhibits
a relatively higher mobility of 30\,$\rm cm^2/Vs$ at room temperature than other perovskites.
This originates in the broad conduction band consisting of Ta 5d.\cite{ktao3_mobility}
Moreover,
$\rm KTaO_3$ is commonly used as a substrate for oxide thin film growth,
\cite{ktao3_surface_trough,ktao3_film2,ktao3_surface_flat}
and as such,
surface treatments of the $\rm KTaO_3$ single-crystal substrate has been extensively studied in the past.
Therefore, once we fabricate an FET on $\rm KTaO_3$, we can expect to apply the method to a number of other functional oxides
integrated on top of $\rm KTaO_3$.

Recently, we have succeeded in fabricating  FET devices on the (001) surface of $\rm SrTiO_3$ single crystals,
and have demonstrated a clear n-channel action.\cite{sto_thesis}
The gentle sputter-deposition of an amorphous $\rm Al_2O_3$ gate insulator
enabled us to avoid deterioration of the interface substantially.
We tried to extend this technique to the "pretreated" surface of $\rm KTaO_3$ single crystals, and successfully fabricated FET devices with distinguished characteristics.
By following a simple pretreatment procedure of the surface,the FET characteristics became quite reproducible.
These $\rm Al_2O_3 / KTaO_3$ FETs show an n-type behavior with an ON/OFF ratio of $10^4$
(at a drain voltage $V_{\rm DS}=1\, V$) and a field effect mobility $\mu_{\rm FE}$ of $\rm 0.4\,cm^2/Vs$ at room temperature.
These values are much better than those of the $\rm Al_2O_3 / SrTiO_3$ FET reported previously.\cite{sto_thesis}

Single crystal substrates of $\rm KTaO_3$ with dimensions of $\rm 10 \times 10 \times 0.5\,mm$ were used.
The surface was polished by the vendor. \cite{earth_seiyaku}
Before the fabrication of devices, the substrate was annealed at high temperatures, as described later.
The structure and fabrication of the device were essentially the same as those in our $\rm Al_2O_3 / SrTiO_3$ FET(Ref.[4]).
Al metal with a thickness of 20\,nm, which forms the source and drain electrodes, was thermally evaporated
onto the (001) surface of $\rm KTaO_3$ through a Ni stencil mask.
Then, an insulating layer of amorphous $\rm Al_2O_3$ with a thickness of 50\,nm
was deposited by a radio-frequency magnetron sputtering at a deposition rate of $\rm 2.5\,\AA/min$.
Finally, a gold wire was attached on the top of the insulating layer by conducting gold paint.
In overhead view, the gold paint covers the entire channel region to act as a gate electrode.
The channel length $L$ of the device was 100\,$\rm \mu m$ and the width $W = $ 400\,$\rm \mu m$.
All the measurements were carried out using an Agilent Technologies 4155C semiconductor parametric analyzer.

Figure 1\,(a) shows the drain-source voltage\,($V_{\rm DS}$)-current\,($I_{\rm DS}$) curves
for various gate voltages $V_{\rm GS}$ of the device fabricated on the $\rm KTaO_3$ single crystal\,( see the discussion of the surface treatment below).
The data were taken at room temperature.
The device performed as an n-type and accumulation-type FET.
By applying positive $V_{\rm GS}$, $I_{\rm DS}$ is greatly enhanced,
whilst for negative $V_{\rm GS}$, no enhancement of  $I_{\rm DS}$ was observed until the gate breakdown.
A clear saturation of $I_{\rm DS}$ (pinch-off) was observed at high $V_{\rm DS}$.
It is worth mentioning that the ON/OFF ratio, defined as the ratio between $I_{\rm DS}$ for $V_{\rm GS}$ = 0\,V
and 5\,V, exceeds $\rm 10^4$ for $V_{\rm DS}$ = 1\,V.
This is two orders of magnitude larger than the best value achieved to date with $\rm SrTiO_3$ FET.
Using the general definition,
\[\mu_{\rm FE} \equiv \frac{\partial I_{\rm DS}}{\partial V_{\rm GS}}\left(\frac{L}{C_iWV_{\rm DS}}\right),\ \ \ \]
we obtain $\mu_{\rm FE} =$  $\rm 0.4\,cm^2/Vs$ at $V_{\rm GS}$ = 5\,V from
the $I_{\rm DS}$-$V_{\rm GS}$ curve for $V_{\rm DS}$ = 1\,V shown in Fig. 3. Here, $C_i = 0.16\,\mu\rm F/cm^2$ is the capacitance per unit area of the $\rm Al_2O_3$ gate insulator.
The deduced value of $\mu_{\rm FE}$ is much higher than the best value of our $\rm Al_2O_3 / SrTiO_3$ FET.

It should be noted that not only the performance of the device but also the reproducibility was much better in $\rm Al_2O_3 / KTaO_3$ FET
than in $\rm Al_2O_3 / SrTiO_3$ FET.
This is because a high-quality surface can be easily obtained on $\rm KTaO_3$ single crystals by a simple heat treatment.
We speculate that the lower sublimation energy of the $\rm KTaO_3$ surface than is reported with the $\rm SrTiO_3$ surface plays an important role in preparing high-quality surfaces in the annealing process.
Figure 2 demonstrates the surface morphology of the $\rm KTaO_3$ single crystal with various heat treatments  as observed by atomic force microscopy (AFM).
In a sample with no heat treatment (``virgin sample''), no structure is observed as seen in Fig.\,2(a).
Annealing at low temperature below 650\,$\rm{}^\circ C$ for one hour in flowing pure oxygen of 50\,sccm at a total pressure of 1atm resulted in the formation of small islands and shallow holes (Fig.\,2(b)).
When annealed at 700\,$\rm{}^\circ C$, we observed flat terraces and steps of single unit-cell height as shown in Fig. 2\,(c).
This so-called step-and-terrace structure is viewed as an evidence for high quality of  surface.
High temperature annealing above 750\,$\rm{}^\circ C$ resulted in the disappearance of step-and-terrace structure,
as well as in the formation of plate-like grains.
A clear FET performance comparable to those in Fig. 1 was reproducibly observed only in devices fabricated on the step-and-terrace surface.

The relatively high mobility of the $\rm Al_2O_3 / KTaO_3$ n-channel FET motivated us to study the field effect at low temperatures.
To see the temperature dependence of $\mu_{\rm FE}$,
the $I_{\rm DS}$-$V_{\rm GS}$ characteristics for $V_{\rm DS}$ = 1\,V were measured with decreasing temperature from 270\,K to 200\,K.
For the data in Fig.3\,(a), we use the same device as used for the data in Fig.1.
The maximum limit of the applied gate voltage at each temperature was such that the gate leakage current could not exceed 10\,pico-amperes.
In Fig.\,3(a), it is clear that the slope of the $I_{\rm DS}-V_{\rm GS}$ curve for a given $I_{\rm DS}$ is
almost independent of temperature.
Shifting each $I_{\rm DS}$-$V_{\rm GS}$ curves by $V_{\rm GS}^* \equiv V_{\rm GS}(I_{\rm DS}= {\rm 100\,nA})$, as seen in the inset of Fig.\,3(a),
all the $I_{\rm DS}$-$V_{\rm GS}$ curves coincide well with each other,
indicating that the apparent temperature dependence of the observed $I_{\rm DS}$-$V_{\rm DS}$ curves has been essentially caused by the temperature dependence of $V_{\rm GS}^*$.
This temperature dependence of $V_{\rm GS}^*$ has been plotted in the upper panel of Fig.\,3(b).
The origin of this strong temperature dependence is as yet unclear.\cite{organic_vth}

The lower panel of Fig.\,3(b) shows $\mu_{\rm FE}$ as a function of temperature.
Since the mobile carrier density at the interface is in general proportional to an ``effective'' gate voltage
$V_{\rm GS}^{\rm eff}\equiv V_{\rm GS}-V_{\rm GS}^*$,
$\mu_{FE}$ at several values of $V_{\rm GS}^{\rm eff}$ are plotted.
All the $\mu_{\rm FE}$ are almost independent of temperature at least down to 200 K.
Indeed, we found that the $I_{\rm DS}$-$V_{\rm DS}$ curves at 300K and 250K for the same $V_{\rm GS}^{\rm eff}$ are almost identical,
confirming the temperature independence of $\mu_{\rm FE}$.
We also note that $\mu_{\rm FE}$ shows an increase with increasing $V_{\rm GS}^{\rm eff}$,
which might be caused by a rapid increase in the mobile carrier density at the interface. 

These results indicate the formation of metallic channel layers at high enough gate voltages
and thus the possibility of superconductivity at cryogenic temperature as seen in bulk $\rm SrTiO_3$ is worth exploring further.
Unfortunately, $I_{\rm DS}$ becomes diminishingly small below a characteristic temperature $T^*$ ($ \simeq\rm 200K$),
and we could not estimate $\mu_{\rm FE}$ below $T^*$.
This is because with decreasing temperature below $T^*$, $V_{\rm GS}^*$ becomes so large that we cannot apply $V_{\rm GS}$ near $V_{\rm GS}^*$
without a risk of destructive breakdown of the gate insulator.
Even in this temperature range, however, when a critically large $\rm V_{\rm GS}$ is applied,
we observe a nonlinear increase of $I_{\rm DS}$ upon increasing $V_{\rm DS}$
and thus enabling $I_{\rm DS}$ to be detected, as seen in the $I_{\rm DS}$-$V_{\rm DS}$ curves in Fig.\,1(c).
This non-linear increase of $I_{\rm DS}$ as a function of $V_{\rm DS}$ indicates
that the current injection to the channel is becoming blocked at lower temperature,
presumably due to the formation of a Schottky barrier at the interface between $\rm KTaO_3$ and the Al electrodes.

In summary, we have demonstrated that $\rm KTaO_3$ is one of the most promising materials for perovskite FET technology.
The FET fabricated on the heat-treated surface of a $\rm KTaO_3$ single crystal showed
accumulation-type behavior and reproducible n-channel transistor characteristics
with $\mu_{\rm FE} > $ 0.4\,$\rm cm^2/V s$ and an ON/OFF ratio greater than $\rm 10^4$ at room temperature.
These specifications are substantially better than those of the best $\rm SrTiO_3$ FET reported so far.
We believe this success of the $\rm Al_2O_3 / KTaO_3$ FET strongly inspires future research into oxide electronics.
At the very least, this technique can be easily extended to a group of perovskite-related oxides
with which a variety of doping induced phenomena possibly give rise to novel functional devices.
Moreover,the temperature-independent $\mu_{\rm FE}$ suggests that
by suppressing the increase of $\rm V_{\rm GS}^*$ at low temperature unexplored metallic state at the $\rm Al_2O_3 / KTaO_3$ interface
will be investigated possibly in the light of low energy phenomena of carrier-injected $\rm KTaO_3$.

We would like to thank Y.~Ishii, T.~Ito, M.~Kawasaki, H.~Sato and A.~Sawa for helpful discussion,
and N.~E.~Hussey for critical reading of the manuscript.

\newpage
\begin{figure}[ht]
\includegraphics[width=15cm]{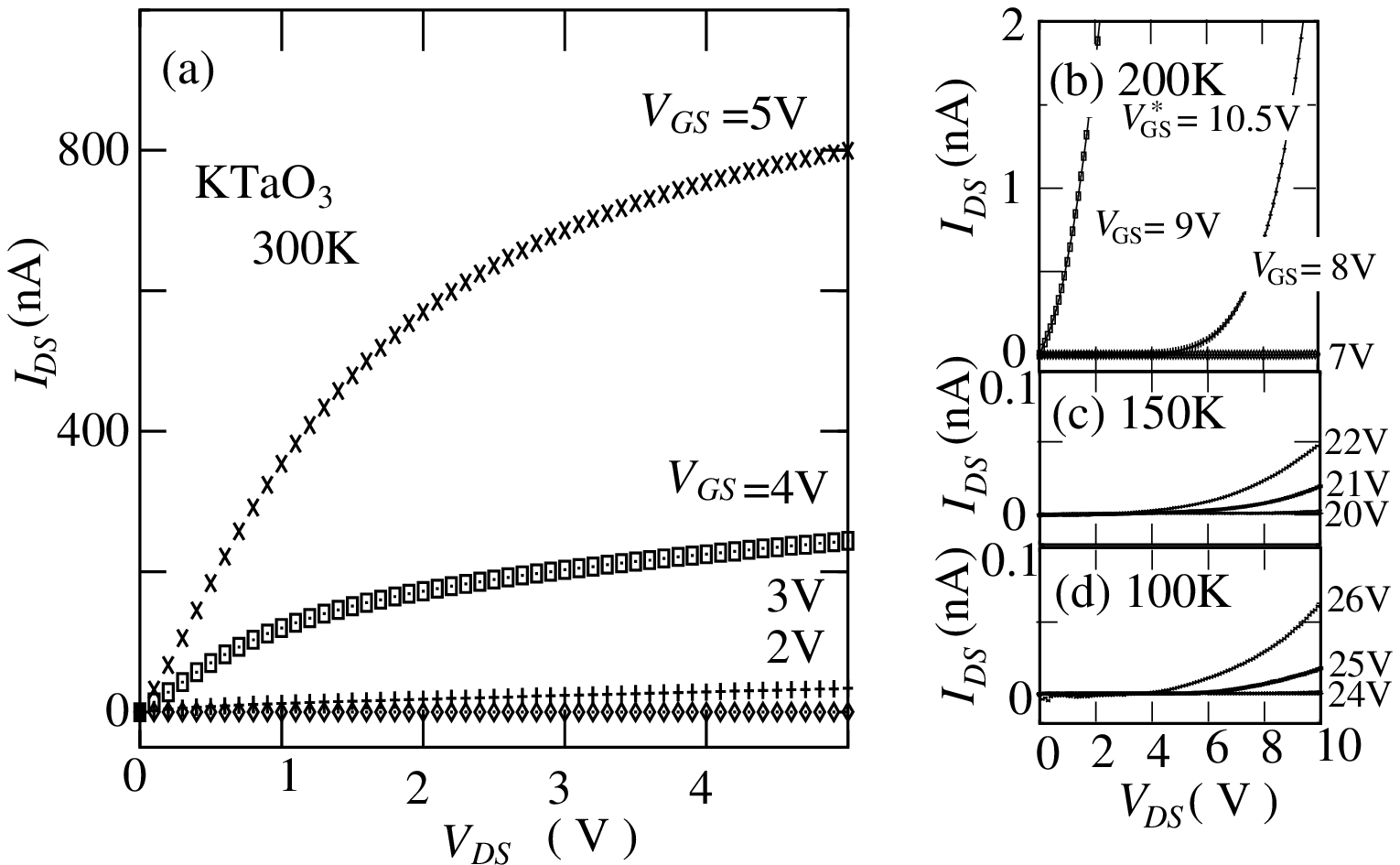}
\caption{
(a) Drain-source current $I_{\rm DS}$ plotted against the drain-source bias $V_{\rm DS}$ of the $\rm Al_2O_3 / KTaO_3$ FET for various gate voltages $V_{\rm GS}$.
The $\rm KTaO_3$ single crystal was annealed at 700\,$\rm{}^\circ C$ prior to the device fabrication.
(b)-(d) $I_{\rm DS}$-$V_{\rm DS}$ plot of the same device at lower temperatures
of (a) 300\,K (b) 200\,K (c) 150\,K (d) 100\,K.
$V_{\rm GS}^*$ at which $I_{\rm DS}$ reached 100\,nA was 10.5\,V at 200K (see text for details).
}
\end{figure}

\begin{figure}[ht]
\includegraphics[width=15cm]{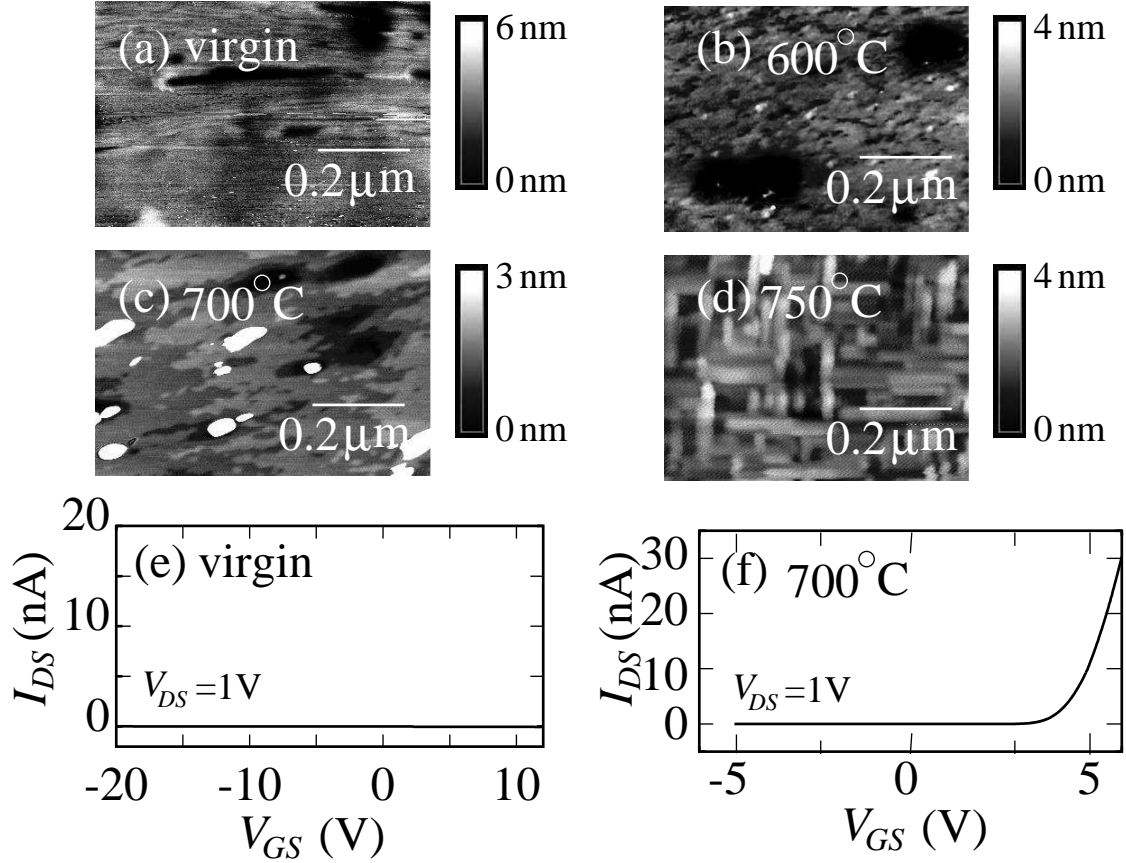}
\caption{
(a)
An atomic force microscopic (AFM) image of the polished (100) surface of a $\rm KTaO_3$ single crystal before annealing.
Neither steps nor terraces were observed.
(b)-(d)
AFM images of the surface of a $\rm KTaO_3$ single crystal annealed at (b) 600, (c) 700, and (d) 750\,$\rm{}^\circ C$.
Flat terraces and steps with a height of one unit cell (0.4\,nm) were observed in Fig.2(c).
(e)
The gate-source bias $V_{\rm GS}$ dependence of the drain-source current $I_{\rm DS}$ for a fixed drain-source bias
$V_{\rm DS}$ = $\rm +1\,V$. The device was fabricated on the polished $\rm KTaO_3$ single crystal.
$I_{\rm DS}$ did not increase by applying gate bias up to the breakdown.
(f)
$V_{\rm GS}$ dependence of $I_{\rm DS}$ for $V_{\rm DS} =$ 1\,V for the device fabricated on the $\rm KTaO_3$ single crystal annealed at 700\,$\rm{}^\circ C$.
Application of a positive gate bias greatly enhances $I_{\rm DS}$.
}
\end{figure}

\begin{figure}[ht]
\includegraphics[width=11cm]{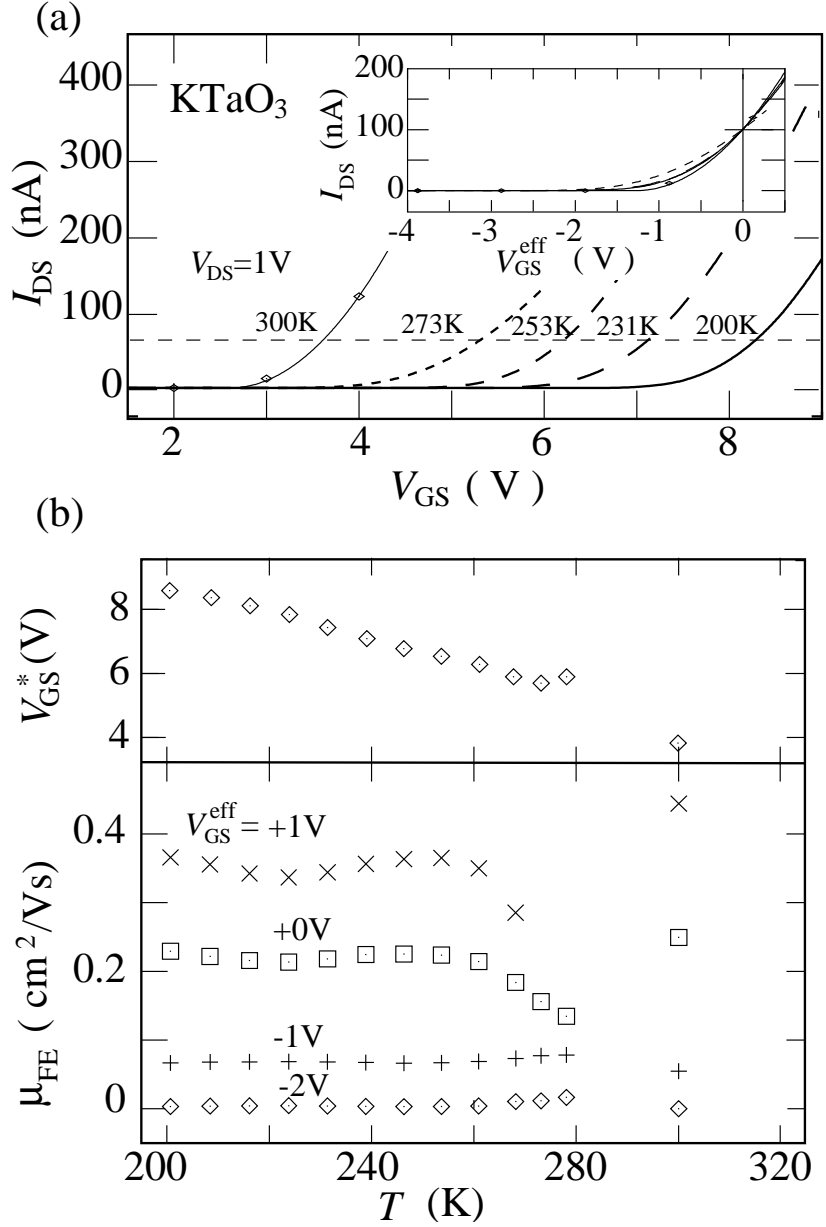}
\caption{
(a) The $V_{\rm GS}$ dependence of $I_{\rm DS}$ for a fixed $V_{\rm DS}$ of $+1\,V$ at various temperatures below 300\,K.
The device was the same one as used in Fig.1\,(a).
The data at 300\,K\,(diamonds) were deduced from Fig.1\,(a) using a  least-squared curve fitting routine (thin line).
The horizontal dashed line corresponds to $I_{\rm DS}$ of 100\,nA.
The inset shows $I_{\rm DS}$ replotted against the effective gate bias $V_{\rm GS}^{\rm eff} \equiv V_{\rm GS}-V_{\rm GS}^*$.
$V_{\rm GS}^*$ is defined as $V_{\rm GS}$ for $I_{\rm DS}$ of 100nA.
The $I_{\rm DS}$-$V_{\rm GS}$ curves at every temperatures are very similar, and thus can hardly be distinguished in this plot.
(b) The upper panel shows $V_{\rm GS}^*$ as a function of the temperature.
The lower panel shows the field-effect mobility $\mu_{\rm FE}$ for $V_{\rm GS}^{\rm eff}$ of -2\,V, -1\,V, 0\,V, and 1\,V
as a function of temperature.
}
\end{figure}
\end{document}